\documentclass[pdflatex,sn-mathphys-num]{sn-jnl}

\usepackage{graphicx}%
\usepackage{multirow}%
\usepackage{amsmath,amssymb,amsfonts}%
\usepackage{amsthm}%
\usepackage{mathrsfs}%
\usepackage[title]{appendix}%
\usepackage{xcolor}%
\usepackage{textcomp}%
\usepackage{manyfoot}%
\usepackage{booktabs}%
\usepackage{algorithm}%
\usepackage{algorithmicx}%
\usepackage{algpseudocode}%
\usepackage{listings}%

\theoremstyle{thmstyleone}%
\theoremstyle{thmstyletwo}%

\theoremstyle{thmstylethree}%

\begin{document}

\title[Article Title]{Enhancing LLM-Based Agents via Global Planning and Hierarchical Execution}


\author[1]{\fnm{Junjie} \sur{Chen}}

\author[1]{\fnm{Haitao} \sur{Li}}

\author[]{\fnm{Jingli} \sur{Yang}}

\author[1]{\fnm{Yiqun} \sur{Liu}}

\author*[1]{\fnm{Qingyao} \sur{Ai}}\email{aiqingyao@gmail.com}

\affil[1]{\orgdiv{DCST}, \orgname{Tsinghua University}, \orgaddress{\city{Beijing}, \postcode{100084}, \country{ China}}}


\abstract{Intelligent agent systems based on Large Language Models (LLMs) have shown great potential in real-world applications. However, existing agent frameworks still face critical limitations in task planning and execution, restricting their effectiveness and generalizability. Specifically, current planning methods often lack clear global goals—leading agents to get stuck in local branches—or produce non-executable plans. Meanwhile, existing execution mechanisms struggle to balance complexity and stability, and their limited action space restricts their ability to handle diverse real-world tasks. To address these limitations, we propose GoalAct, a novel agent framework that introduces a continuously updated \textbf{global planning} mechanism and integrates a \textbf{hierarchical execution} strategy. GoalAct decomposes task execution into high-level skills, including searching, coding, writing and more, thereby reducing planning complexity while enhancing the agents' adaptability across diverse task scenarios. We evaluate GoalAct on LegalAgentBench, a benchmark with multiple types of legal tasks that require the use of multiple types of tools. Experimental results demonstrate that GoalAct achieves state-of-the-art (SOTA) performance, with an average improvement of 12.22\% in success rate. These findings highlight GoalAct's potential to drive the development of more advanced intelligent agent systems, making them more effective across complex real-world applications. Our code can be found at \url{https://github.com/cjj826/GoalAct}.}

\keywords{Large Language Model, Agent, Global Planning, Hierarchical Execution}



\maketitle

\section{Introduction}\label{sec1}

\label{sec:intro}
In recent years, intelligent agent systems based on Large Language Models (LLMs) have shown remarkable potential in practical applications \cite{wang2024survey,lai2024autowebglm}. This advancement comes mainly from two aspects: (1) LLMs have continuously improved their core capabilities, such as instruction-following, tool utilization, programming, writing, and logical reasoning \cite{schick2023toolformer, achiam2023gpt, liu2024deepseek}; (2) the ongoing advancement of agent frameworks (e.g., Plan-and-Solve \cite{wang2023plan}, ReAct \cite{yao2023react}, CodeAct \cite{wang2024executable}) has enabled agents to autonomously perform task planning (what to do) and task execution (how to do), facilitating the achievement of target objectives through interactions with the environment.

Although existing agent frameworks demonstrate a certain level of autonomy in \textbf{task planning and execution}, substantial limitations remain within these two critical stages:

Firstly, at the planning level, current methods either lack clear global goals or generate plans that are hard to execute. Some frameworks, such as ReAct, adopt an incremental reasoning approach of ``Thought-Action-Observation'', focusing only on the immediate step without a comprehensive global perspective. Consequently, these frameworks frequently become stuck in local optima during tasks involving multiple branches. In contrast, methods like Plan-and-Execute \cite{topsakal2023creating} attempt to enhance task-solving capabilities by first generating a global plan and subsequently executing it, while dynamically adjusting the plan based on feedback obtained during execution. However, such methods often fail to effectively integrate concrete executable actions into the global plan, resulting in plans that exceed the agents' action space.

Secondly, at the execution level, existing methods face a trade-off between complexity and stability, while also being constrained by a limited action space. Frameworks like ReAct 
primarily rely on \textbf{text or json formats} to invoke external tools. While these formats offer a straightforward mechanism for tool interaction, they lack the capability to handle complex logic structures, such as loops and conditional branches. CodeAct attempts to unify the agent's action space using \textbf{python code}, allowing for more sophisticated tool invocation logic. However, despite its greater expressive power, it increases execution complexity. In real-world scenarios where tool invocation outcomes are inherently unpredictable, highly complex invocation logic is more prone to errors, making the system less stable. Moreover, not all tasks can be effectively solved using code alone, such as advanced writing tasks (e.g., legal document generation \cite{li2025casegen}) and logical/mathematical reasoning (e.g., Lateral Thinking Puzzles \cite{liuagentbench}).

These limitations lead to discrepancies between the agents' intended decisions and their actual executed behaviors, thereby constraining both the effectiveness and generalizability of these agents when addressing complex real-world tasks. To address the above issues, we propose the \textbf{GoalAct} framework, aiming to enhance LLM-based agents via global planning and hierarchical execution:

\begin{itemize}
    \item \textbf{Global planning:} we introduce a continuously updatable global planning mechanism that tightly couples planning and execution, enabling agents to establish clearer long-term goals while ensuring the feasibility of plans.
    \item \textbf{Hierarchical execution:} we argue that the complexity of real-world actions cannot be effectively managed by \textbf{existing single-stage execution methods}, which require agents to simultaneously determine the appropriate skills or methods, select relevant tools, and configure their parameters. Inspired by human practice—first identifying high-level skills, then selecting suitable tools, and finally refining execution details—we propose a hierarchical execution framework that decomposes task execution into distinct high-level skills, such as searching, coding, writing, reasoning and more. This hierarchical structure offers two key advantages: (1) it significantly simplifies global planning, as the plan only needs to specify appropriate high-level skills and their objectives rather than low-level details; and (2) it is inherently scalable, enabling the dynamic addition and selection of skills to flexibly adapt to diverse and evolving task scenarios.
\end{itemize}

To evaluate the effectiveness of GoalAct, we conducted experiments using LegalAgentBench \cite{li2024legalagentbench}, a novel benchmark that poses no risk of data leakage and requires external tool invocation and legal knowledge for task completion. Experimental results demonstrate that GoalAct achieves state-of-the-art (SOTA) performance, with an average improvement of 12.22\% in success rate.

\section{Related Work}
\subsection{Large Language Models}
Large Language Models (LLMs) have experienced rapid development in recent years, significantly advancing natural language understanding and generation capabilities. Models like ChatGLM \cite{glm2024chatglm,zeng2022glm}, Qwen \cite{bai2023qwen} and GPT \cite{achiam2023gpt} demonstrate remarkable abilities in language comprehension, multi-turn dialogue management, and instruction following \citep{radford2019language,achiam2023gpt}. Techniques such as contextual learning, in-context instruction tuning, and chain-of-thought reasoning have enabled LLMs to handle complex reasoning tasks more effectively \citep{ouyang2022training, wei2021finetuned, wei2022chain}. These advances have led to LLMs being integrated into various downstream applications, from knowledge-based question answering to tool-use scenarios, thereby driving the development of more sophisticated intelligent agents.

\subsection{LLM-Based Agents}
Leveraging the advancements of LLMs, researchers have built agents capable of reasoning, planning, and interacting within complex environments. Approaches such as ReAct \cite{yao2023react} combine reasoning and acting to allow agents to interact iteratively with external tools, enhancing the agent's factuality and problem-solving capabilities. Voyager \cite{wang2023voyager} introduces code-based action execution, enabling an LLM agent to operate effectively within dynamic game environments by generating executable python code to navigate and interact within Minecraft. Similarly, TaskWeaver \cite{qiao2023taskweaver} translates user instructions into executable code for efficient task completion in structured data analysis environments. Other frameworks like CodeAct \cite{wang2024executable} emphasize tool integration, enabling agents to leverage external coding environments for complex problem-solving tasks, significantly enhancing their operational flexibility. Despite these promising developments, current LLM-based agent frameworks still face limitations in task planning and execution, highlighting ongoing research opportunities.
\section{Methodology}
\begin{figure}[htbp]
    \centering
    \includegraphics[width=\textwidth]{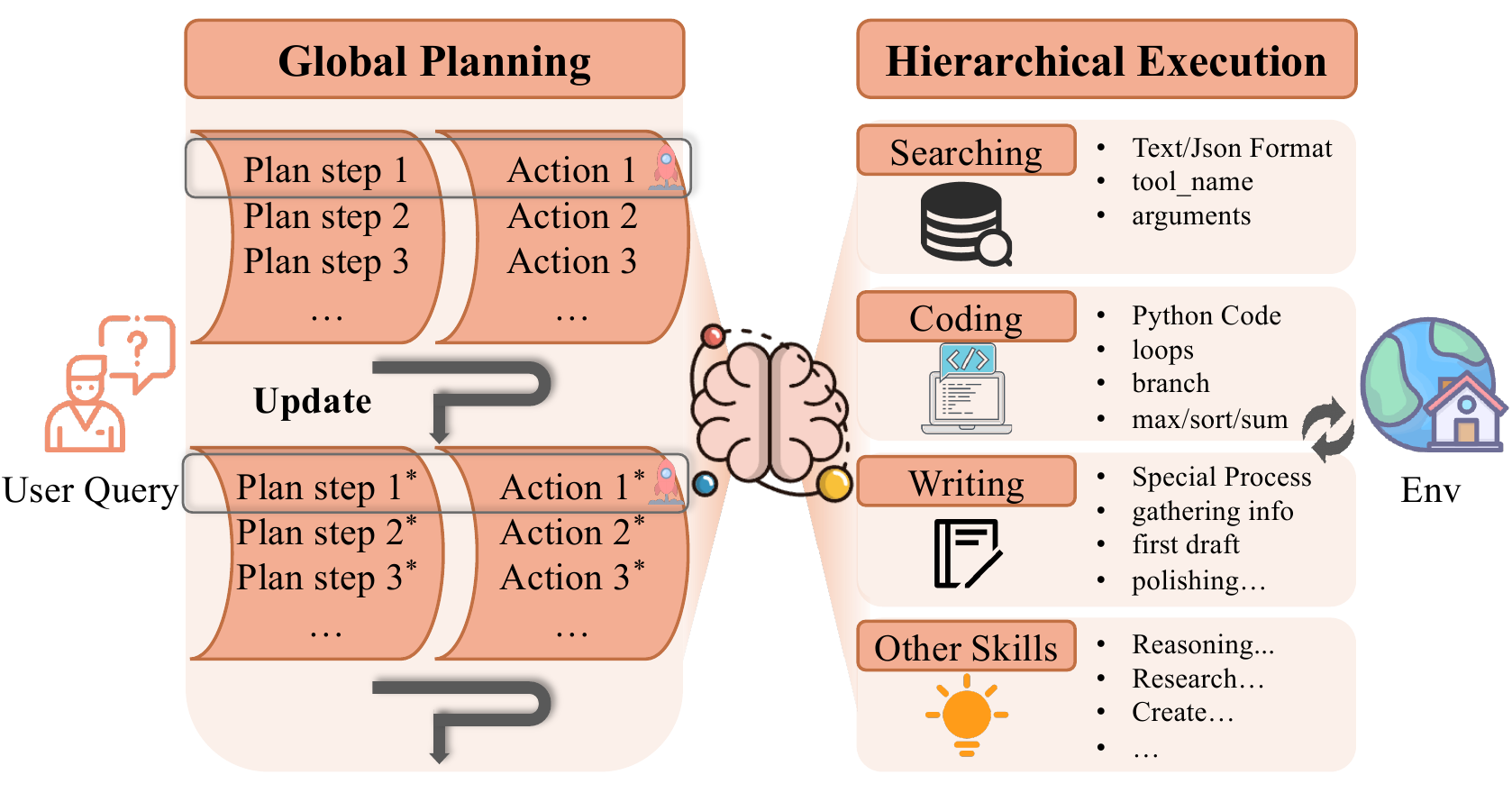} 
    \caption{The framework of our GoalAct. It emphasizes continuously updatable \textbf{ global planning}, which harmonizes the clarity of long-term goals with actionable steps, and \textbf{hierarchical execution}, which decomposes task execution into high-level skills. }
    \label{fig:goalact}
\end{figure}

Figure \ref{fig:goalact} shows the framework of our GoalAct. Next, we introduce it in detail.

\subsection{Global Planning}
In our GoalAct framework, we continuously maintain and update a global plan $G$, defined as Equation~\eqref{eq:global_plan} shows:  

\begin{equation}
G = (P_1 A_1, P_2 A_2, P_3 A_3, \dots, P_n A_n)
\label{eq:global_plan}
\end{equation}
Here, \( P_i \) denotes the \( i \)-th plan step, and \( A_i \) represents its corresponding action. Since providing detailed action descriptions directly within the global plan is challenging, we only require the specification of high-level skills, which will be introduced next in the section \ref{Hierarchical Execution}. The final step, \( A_n \), is always designated as $Finish$.

At each time step \( t \), the agent updates \( G \) according to Equation~\eqref{eq:plan_update}:  
\begin{equation}
G_t = \pi (Q \mid T \mid S_t)
\label{eq:plan_update}
\end{equation}  
where \( \pi \) is the update policy (detailed in the Table \ref{prompt:GP}), \( Q \) is the user query, \( T \) is the set of available tools, and \( S_t \) is the historical record at time \( t \), as Equation~\eqref{eq:history} shows:  
\begin{equation}
S_t = (P_1 A_1 O_1, P_2 A_2 O_2, \dots, P_{t-1} A_{t-1} O_{t-1})
\label{eq:history}
\end{equation}  
Here, \( O_i \) represents the observation from the execution of \( A_i \). When \( t = 1 \), \( S_t \) is empty.

Compared to existing methods, GoalAct tightly integrates global planning with execution through the continuously updated global plan. By dynamically adjusting the plan based on real-time feedback, it ensures both a coherent overall strategy and practical executability, ultimately enhancing the agents across diverse scenarios.

\subsection{Hierarchical Execution}
\label{Hierarchical Execution}
To reduce the complexity of considering executable actions within global planning and to enable the agent to flexibly adapt its action complexity based on task difficulty, we introduce a hierarchical execution framework. As stated in section \ref{sec:intro}, we argue that existing single-stage execution methods struggle to manage real-world action complexity, as they require agents to simultaneously determine appropriate skills or techniques, select tools and configure them. Inspired by human action processes—first identifying necessary skills, then selecting appropriate tools, and finally fine-tuning execution parameters—our framework firstly structures execution into distinct skills, allowing global planning to focus on high-level skills rather than low-level details, while maintaining flexibility and extensibility, thereby ensuring a well-organized execution process and enhancing the agents. Next, we will introduce several representative skills:

\textbf{Searching}: This skill primarily functions as a tool invocation mechanism for information retrieval, utilizing a text or json formats similar to the ReAct. Its key advantage lies in its simplicity and stability in complex external environments, where simpler probing strategies tend to be more effective, as they facilitate error analysis and behavioral refinement. However, its major limitation is its restricted expressiveness, as it cannot handle complex tool calls or data processing logic, such as conditional branches and loops. Nevertheless, this approach remains valuable in some simpler scenarios.

\textbf{Coding}: This skill leverages python code to invoke tools and process data. Its strength lies in its ability to efficiently implement complex logic, including branching and looping, thereby expanding the agent’s action space and enhancing execution efficiency. However, this increased action space comes at the cost of greater complexity. While ongoing research \cite{wang2023voyager} aims to enable agents to generate more robust code, external environments remain inherently unpredictable, and even the most sophisticated code can produce unforeseen bugs. In such cases, a simpler Searching approach often serves as a more effective exploration method.

\textbf{Writing}: In real-world applications, many specialized tasks , which involve specific and intricate logic, cannot be effectively addressed through either Searching or Coding. For example, writing legal documents typically requires gathering legal articles, legal knowledge, and similar cases, followed by composing a preliminary draft and finally structuring the output according to specific format requirements. In such scenarios, although coding excels in handling complex branching and conditional logic, it does not directly contribute to completing the task.

Beyond these skills, expanding the skill pack according to task-specific requirements is essential. For instance, reasoning tasks may require specialized reasoning skills, scientific research tasks may demand research skills, and artistic creation tasks may call for creative skills. Notably, this expansion process is easy and highly iterative, enabling continuous adaptation and refinement to meet evolving task demands.

\section{Experiment}
\subsection{Experimental Setup}
\subsubsection{Tasks Selection}
To validate the effectiveness of GoalAct, it is essential to ensure the reliability of the evaluation benchmark. However, existing well-known datasets may have already been used in training, posing a risk of data leakage. Moreover, we aim to assess the agent’s performance in an unfamiliar environment. Therefore, we select \textbf{LegalAgentBench} \cite{li2024legalagentbench} as our benchmark. As a newly released dataset, it eliminates the risk of data leakage while covering a comprehensive range of task difficulties. Additionally, solving its tasks requires external data and legal knowledge, making it a suitable and robust test for the capabilities of agents.

The data in LegalAgentBench has been meticulously verified by human experts and includes a total of \textbf{300 tasks across six different types}. Table \ref{table:statistics} presents the detailed statistics of these tasks. Among them, \textbf{1-hop to 5-hop} tasks represent logical reasoning problems with varying solution path lengths, where a higher hop count indicates increased task complexity. Beyond logical reasoning tasks, the dataset also features a \textbf{writing} category, specifically \textbf{writing a defense document}. In this scenario, the agent must query basic information about the plaintiff, defendant, and their lawyer, while simultaneously retrieving relevant legal knowledge and relevant articles to construct a defense against the complaint.

\begin{table}[t]
\caption{Detailed Statistics of LegalAgentBench Tasks.}
\centering
{%
\begin{tabular}{cccccccc}
\hline
Attribute                  & 1-hop & 2-hop & 3-hop & 4-hop  & 5-hop  & Writing & ALL    \\ \hline
\# Task & 80 & 80 & 60 & 40 & 20 & 20 & 300 \\

Avg. length per query   & 88.29 & 87.90 & 99.37 & 118.33 & 110.25 & 1059.95 & 160.65 \\
Avg. length per answer     & 74.20 & 40.84 & 45.53 & 63.48  & 86.20  & 678.75  & 99.24  \\
\# Avg. key\_answer per query & 1.88  & 1.44  & 1.20  & 1.40   & 2.25   & 10.25   & 2.14   \\
Avg. length of key\_answer & 10.59 & 5.94 & 6.07 &6.59 &6.93 &12.58 &9.28 \\
\# Avg. key\_middle per query & 0.13  & 1.45  & 2.87  & 4.78   & 5.60   & 6.20    & 2.42   \\
Avg. length of key\_middle     & 9.20 &9.72 &10.95 &11.35 &11.25 &7.21 &10.24   \\
 \hline
\end{tabular}
}
\label{table:statistics}
\end{table}

\subsubsection{Baselines}
We evaluated three well-known LLMs on LegalAgentBench: GLM-4-Plus~\cite{glm2024chatglm}, Qwen-max~\cite{bai2023qwen}, GPT-4o-mini~\cite{hurst2024gpt} (gpt-4o-mini-2024-07-18). All LLMs are evaluated through API calls. To ensure the reproducibility of the results, we set the temperature to 0.

For each LLM, we implemented four different methods. (1) \textbf{Plan-and-Solve}: Outline a complete plan and execute it step by step. (2) \textbf{Plan-and-Execute}: Develop a multi-step plan and complete it sequentially. After completing a task, the LLM can reassess the plan and make appropriate adjustments. (3) \textbf{ReAct}: Perform reasoning incrementally through the ``thought-action-observation'' process, integrating reasoning and tool usage. (4) \textbf{CodeAct}: Utilize Python code as the agent's action space, enabling the invocation of multiple tools within a single execution.

When given a task, the model first determines which tools are needed, and then uses the selected tools to gradually solve the task. 
When the LLM outputs \textit{Finish} or reaches the maximum iteration limit $T=10$, it summarizes the current trajectory and provides the final answer. We include two examples for each process to guide the model in using the tools and following the specified output format.

\subsubsection{Metrics}
We use \textbf{Success Rate} as the metric, which measures the proportion of key\_answer elements included in the LLM's output. Assume a dataset $\mathcal{D}$ consisting of $N$ data points, where each data point includes a keyword set $\mathcal{K}_i$ and a model output $\mathcal{O}_i$. The success rate $s_i$ for the $i$-th data point is computed as Equation \ref{eq:success} shows:  
\begin{equation}
    s_i = \frac{|\mathcal{M}_i|}{|\mathcal{K}_i|}
\label{eq:success}
\end{equation}
where $\mathcal{M}_i = \{k \in \mathcal{K}_i \mid k ~\text{appears in} ~\mathcal{O}_i \}$. The notation $|\cdot|$ represents the number of elements in a set. We report the average success rate across all tasks in the results.

\begin{table}[]
\caption{The success rate of different methods on LegalAgentBench. P-S represents the Plan-and-Solve method,
and P-E represents the Plan-and-Execute method. The best results are highlighted in bold.}
\label{tab:my-table1}
\begin{tabular}{lllllllll}
\hline
Model                        & Method           & 1-hop           & 2-hop           & 3-hop           & 4-hop           & 5-hop           & Writing         & ALL             \\ \hline
\multirow{5}{*}{GPT-4o-mini} & P-S              & 0.7117          & 0.3375          & 0.2750          & 0.2583          & 0.1250          & 0.6314          & 0.4196          \\
                             & P-E              & 0.7444          & 0.3771          & 0.3250          & 0.2417          & 0.1417          & 0.6681          & 0.4503          \\
                             & ReAct            & 0.9333          & 0.6500          & 0.4000          & 0.4208          & 0.2583          & 0.6087          & 0.6161          \\
                             & CodeAct          & 0.9058          & 0.7938          & 0.4167          & 0.3875          & 0.1683          & 0.4202          & 0.6275          \\
                             & \textbf{GoalAct} & \textbf{0.9556} & \textbf{0.8771} & \textbf{0.6250} & \textbf{0.5625} & \textbf{0.4517} & \textbf{0.7981} & \textbf{0.7720} \\ \hline
\multirow{5}{*}{Qwen-max}    & P-S              & 0.8469          & 0.4958          & 0.4083          & 0.3792          & 0.2333          & 0.4836          & 0.5381          \\
                             & P-E              & 0.8594          & 0.5896          & 0.3583          & 0.4083          & 0.3017          & 0.5539          & 0.5695          \\
                             & ReAct            & 0.9062          & 0.7917          & 0.6333          & 0.5833          & 0.6083          & 0.6662          & 0.7422          \\
                             & CodeAct          & 0.8442          & 0.8583          & 0.6333          & 0.7583          & 0.5117          & 0.6522          & 0.7594          \\
                             & \textbf{GoalAct} & \textbf{0.9531} & \textbf{0.9062} & \textbf{0.8167} & \textbf{0.7917} & \textbf{0.6117} & \textbf{0.8220} & \textbf{0.8603} \\ \hline
\multirow{5}{*}{GLM-4-Plus}  & P-S              & 0.8519          & 0.4667          & 0.4167          & 0.3583          & 0.1167          & 0.7522          & 0.5406          \\
                             & P-E              & 0.8419          & 0.5000          & 0.3667          & 0.3458          & 0.1167          & 0.7679          & 0.5363          \\
                             & ReAct            & 0.9131          & 0.8104          & 0.6417          & 0.6167          & 0.4300          & 0.7659          & 0.7499          \\
                             & CodeAct          & 0.9221          & 0.7812          & 0.4167          & 0.5458          & 0.2333          & 0.5836          & 0.6648          \\
                             & \textbf{GoalAct} & \textbf{0.9506} & \textbf{0.9187} & \textbf{0.8500} & \textbf{0.7375} & \textbf{0.6983} & \textbf{0.8643} & \textbf{0.8710} \\ \hline
\end{tabular}
\end{table}
\subsection{Experimental Results}
\subsubsection{Main Results}
Table \ref{tab:my-table1} presents the results of different methods on LegalAgentBench, from which we draw the following conclusions:

\begin{itemize}
    \item \textbf{GoalAct achieves SOTA performance across different LLM series and varying task difficulties.} Compared to the second-best method, GoalAct improves average performance by \textbf{14.45\% on GPT-4o-mini}, \textbf{10.09\% on Qwen-max}, and \textbf{12.11\% on GLM-4-Plus}, with an overall average improvement of \textbf{12.22\%}. These results demonstrate the broad effectiveness of GoalAct.

    \item \textbf{The inherent limitations of various baseline methods undermine their performance.}  
\textbf{P-S} executes a static global plan, which limits its adaptability in complex scenarios, leading to poor performance.
\textbf{P-E} dynamically updates its global plan, demonstrating improvements over P-S in some scenarios. However, as its global plan does not account for task execution, the execution feasibility of its plan in some tasks is weak, ultimately resulting in performance degradation.
\textbf{ReAct} exhibits relatively strong performance, but the absence of global planning and its reliance on action formats constrained by text or json impose an upper bound on its effectiveness in certain scenarios.  
\textbf{CodeAct} extends the action space through python code; however, in complex scenarios, invoking multiple tools within a single code execution often results in errors. Moreover, experimental results show that CodeAct performs poorly in Writing tasks, highlighting its significant limitations.

    \item \textbf{As the task difficulty increases, the performance gain of GoalAct over the second-best method becomes more pronounced.}  
    Specifically, the relative improvements on average are \textbf{3.26\% for 1-hop}, \textbf{7.98\% for 2-hop}, \textbf{20\% for 3-hop}, \textbf{9.86\% for 4-hop}, and \textbf{15.5\% for 5-hop}. Furthermore, in Writing tasks, GoalAct achieves a 12.74\% improvement, further validating its strong adaptability and generalization across different task types.
\end{itemize}

\subsubsection{Ablation Study}
\begin{table}[]
\caption{An ablation study was conducted to evaluate the effectiveness of GoalAct with the GLM-4-Plus as the base LLM. The notation `w/o' indicates experiments where specific modules were removed.}
\label{tab:my-table2}
\begin{tabular}{llllllll}
\hline
Method           & 1-hop           & 2-hop           & 3-hop           & 4-hop           & 5-hop           & Writing         & ALL             \\ \hline
\textbf{GoalAct} & \textbf{0.9506} & \textbf{0.9187} & \textbf{0.8500} & \textbf{0.7375} & \textbf{0.6983} & \textbf{0.8643} & \textbf{0.8710} \\
w/o global plan  & 0.9281          & 0.7375          & 0.8000          & 0.6208          & 0.6950          & 0.8454          & 0.7896          \\
w/o searching    & 0.8950          & 0.8063          & 0.7500          & 0.6708          & 0.6100          & 0.8521          & 0.7906          \\
w/o coding       & 0.8381          & 0.7833          & 0.6333          & 0.6208          & 0.4833          & 0.8447          & 0.7304          \\
w/o writing      & 0.9381          & 0.9000          & 0.7667          & 0.6708          & 0.5767          & 0.8247          & 0.8264          \\ \hline
\end{tabular}
\end{table}

To evaluate the effectiveness of \textbf{Global Planning and Hierarchical Execution}, we conducted an ablation study by systematically removing key components of GoalAct with GLM-4-Plus as the base LLM. Table \ref{tab:my-table2} presents the results, from which we observe the following: Removing the \textbf{global plan} reduces average performance by \textbf{8.14\%}, \textbf{8.04\%} for \textbf{Searching}, \textbf{14.06\%} for \textbf{Coding}, and \textbf{4.46\%} for \textbf{Writing} (with Writing tasks specifically declining by \textbf{3.96\%}). These ablation results validate the rationality of GoalAct’s design and highlight the necessity of synergy among its components to achieve optimal performance.

\subsubsection{Case Study}
\begin{figure}[h]
    \centering
    \includegraphics[width=\textwidth]{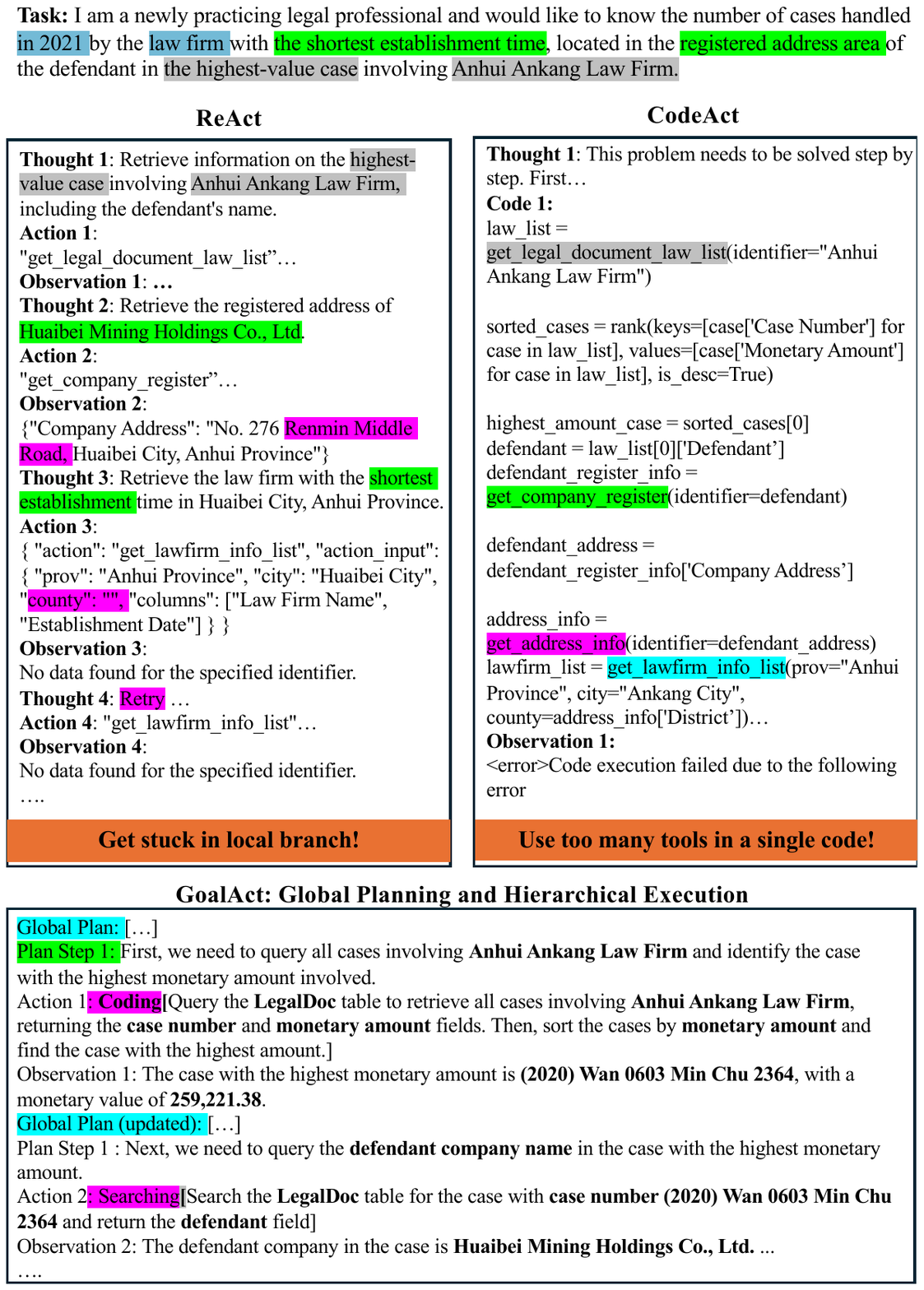} 
    \caption{A specific example of ReAct, CodeAct and GoalAct in our experiments.}
    \label{fig:case}
\end{figure}

Figure \ref{fig:case} presents a specific example of the competitive agent frameworks in our experiment, including ReAct, CodeAct, and GoalAct, with GLM-4-Plus as the base LLM. We observe that ReAct has restricted action space due to the text or json formats and its tendency to repeatedly attempt solutions, often getting stuck in local branches. CodeAct leverages code to express complex logic, which enhances its action space. However, its approach of invoking multiple tools within a single piece of code introduces execution difficulties, as each tool’s output may contain uncertainties. In contrast, our proposed method, GoalAct, effectively avoids the local branch issue by implementing a global planning strategy. Additionally, its hierarchical execution mechanism enables the agent to flexibly choose appropriate skills based on task complexity, utilizing searching for simpler tasks and coding for more complex ones. These observations demonstrate that GoalAct offers superior performance and greater generalizability.

\section{Conclusion}
This paper presents GoalAct, which enhances LLM-based agents by integrating global planning and hierarchical execution. Experimental results on the LegalAgentBench benchmark demonstrate that GoalAct achieves state-of-the-art (SOTA) performance, with an average improvement of 12.22\% over existing methods. In the future, GoalAct can be integrated with mechanisms such as agent reflection~\cite{shinn2023reflexion} and memory~\cite{zhong2024memorybank} to facilitate the development of more advanced and intelligent agent systems.

\bibliography{sn-bibliography}
\clearpage
\section*{Appendix}
\begin{table*}[h]\centering
\caption{The prompt used in the GoalAct for global planning.}
\resizebox{\textwidth}{!}{%
\begin{tabular}{l}
\hline
\textbf{The prompt used in the GoalAct for global planning.} \\ \midrule
\begin{tabular}[c]{@{\ }p{\textwidth}@{\ }}You are a planner skilled in solving complex tasks. You use the three alternating steps of ``Thinking, Acting, and Observing" to solve question-answer tasks based on the provided data tables.

Thinking involves reasoning about the current situation and determining the next subproblem to solve the current issue.

Acting is performing an operation based on the results of your thinking. It must be one of the following four types: \\1. Searching: Retrieve a record from a data table based on information, or filter multiple records that meet specific attribute values.
\\2. Coding: If the problem is too complex to be solved by querying alone, you may attempt to program a solution. You can filter, sort, sum, or iterate over queried data.
\\3. Writing: If content generation (such as a defense statement) is needed, you should attempt writing to resolve the issue.
\\4. Finish: Provide the final answer and terminate the task.

The action must be detailed within square brackets [] and should accurately identify the correct table and returned fields.

Observing is the information obtained after an action. If all observations sufficiently answer the problem, provide the final response and end the task.

You should take the necessary steps. Ensure that your response strictly follows the format above. Specifically, the action must be one of the listed types, and all actions should terminate with an end.

Problem to be solved:
\{question\}

The data tables involved include:
\{table\_used\_prompt\}

Available API\_tool:
\{tool\_prompt\}

Here are some reference examples:
\{memory\}

Reference examples end.

Question: \{question\}

Existing planning chain:
\{scratchpad\}

Please continue to think and execute based on the existing planning chain logic. Ensure that you avoid repeating any previous thought paths or actions already taken.
If the current planning chain sufficiently resolves the issue, directly output the action as ``Finish"

Please output the result in the following JSON format, which can be parsed using Python's json.loads function. Provide only the problem decomposition result without explanation or direct answers:

``json
[
    \{\{
        ``Thinking'': ``'',
        ``Action'': ``''
    \}\}
]
''\end{tabular} \\ \hline
\end{tabular}%
}
\label{prompt:GP}
\end{table*}

\end{document}